\documentclass[12pt]{article}
\pdfoutput=1
\usepackage{ifpdf}
\usepackage[a4paper,top=2.6cm,width=17cm, bottom=2.6cm, left=2cm]{geometry}
\usepackage{fancyhdr}
\usepackage{microtype}
\usepackage[utf8]{inputenc}
\usepackage{mathdots}
\usepackage{setspace}
\usepackage{cite}
\usepackage{hyperref}
\usepackage{url}
\usepackage{amsmath,amssymb,amscd,braket}
\usepackage{amsfonts}
\usepackage{color}
\usepackage{graphicx}
\usepackage{xcolor}
\usepackage{accents}
\usepackage[indent=30pt]{parskip}
\usepackage{authblk}

\newcommand{\im}{\mathfrak{Im}}

\newcommand{\ww}{\tilde{w}}
\newcommand{\kk}{\tilde{k}}
\DeclareMathOperator\arctanh{arctanh}

\newcommand{\bea}{\begin{eqnarray}}
\newcommand{\eea}{\end{eqnarray}}
\newcommand{\be}{\begin{equation}}
\newcommand{\ee}{\end{equation}}
\newcommand{\ba}{\begin{align}}
\newcommand{\ea}{\end{align}}

\newcommand{\rf}[1]{Eq.~(\ref{#1})}

\begin{document}

\title{
{\bf The Hydrohedron: \\ Bootstrapping Relativistic Hydrodynamics}\\
}

\renewcommand\Authfont{\scshape\small}

\author[1]{Michal P. Heller\thanks{michal.p.heller@ugent.be}}
\author[2]{Alexandre Serantes\thanks{alexandre.serantes@ub.edu}}
\author[3,4]{Micha\l{} Spali\'nski\thanks{michal.spalinski@ncbj.gov.pl}}
\author[5]{Benjamin Withers\thanks{b.s.withers@soton.ac.uk}}

\affil[1]{Department of Physics and Astronomy, Ghent University, 9000 Ghent, Belgium}
\affil[2]{Departament de Física Quàntica i Astrofísica, Institut de Ciències del Cosmos (ICCUB), Facultat de Física, Universitat de Barcelona, Martí i Franquès 1, E-08028 Barcelona, Spain}
\affil[3]{Physics Department, University of Bia\l{}ystok, 15-245 Bia\l{}ystok, Poland}
\affil[4]{National Center for Nuclear Research, 00-681 Warszawa, Poland}
\affil[5]{Mathematical Sciences and STAG Research Centre, University of Southampton, Highfield, Southampton SO17 1BJ, UK}


\date{}

\maketitle

\begin{abstract}

\noindent As an effective theory, relativistic hydrodynamics is fixed by symmetries up to a set of transport coefficients. A lot of effort has been devoted to explicit calculations of these coefficients. Here we propose a shift in perspective: we deploy bootstrap techniques to rule out theories that are inconsistent with microscopic causality. What remains is a universal convex geometry in the space of transport coefficients, which we call \emph{the hydrohedron}. The landscape of all consistent theories necessarily lie inside or on the edges of the hydrohedron. We analytically construct cross-sections of the hydrohedron corresponding to bounds on transport coefficients that appear in sound and diffusion modes for theories without stochastic fluctuations.
\end{abstract}

\section{Introduction}

Hydrodynamics is a universal description of systems tending towards thermal equilibrium. It is formulated as an effective theory, order-by-order in a gradient expansion, which at the classical linearised level can be mapped to the expansion of the mode frequencies in powers of the wave-vector:
\be
\omega(k) = \sum_{n=1}^\infty c_{n} k^{n}.
\label{omega_series}
\ee
Such hydrodynamic Taylor series expansions have been studied in a wide variety of examples and have been found to have a finite radius of convergence \cite{Withers:2018srf,Grozdanov:2019kge,Grozdanov:2019uhi,Abbasi:2020ykq, Jansen:2020hfd,Heller:2020hnq}. The complex $c_n$ are a collection of transport coefficients which include the speed of sound and the diffusion constant. The aim of this paper is to characterise the set of physically acceptable collections of transport coefficients, which can be thought of as the landscape of hydrodynamic theories. We propose to chart its boundaries by imposing the causality condition~\cite{Heller:2022ejw}, 
\be
\boxed{v_{\text{LC}}\, |\im\, k|-\im\,\omega(k) \geq 0.}\label{cond}
\ee
where $v_{\text{LC}}$ is the lightcone speed (we set $v_{LC}=1$ in what follows).
This condition arises axiomatically, a consequence of position space retarded Green's functions being tempered distributions and causality dictating support only in the appropriate light cone. This implies that certain regions of its Fourier transform are analytic, thus restricting where physical modes can appear. 
In Ref.~\cite{Heller:2022ejw} we used \rf{cond} to prove that all dissipative hydrodynamic expansions \eqref{omega_series} have a finite radius of convergence $R$, and establish two-sided bounds on all dimensionless combinations $R^{n-1}c_n$.

In the present work we propose to view \eqref{cond} in a completely new way. 
Taking only the minimal ingredients of analyticity of the mode functions at $k=0$ \eqref{omega_series}, and the causality of the Green's function \eqref{cond}, we seek to constrain where the landscape of admissible transport coefficients lies.
This strategy adheres to the bootstrap approach to theoretical physics problems, which carves out a space of consistent theories using fundamental principles. It also enables us to profit from technologies used in other such programmes such as the modern conformal bootstrap \cite{Rattazzi:2008pe, El-Showk:2012cjh, Simmons-Duffin:2016gjk, Poland:2018epd} and the S-matrix bootstrap (for example \cite{Arkani-Hamed:2020blm, Caron-Huot:2022ugt, Kruczenski:2022lot, deRham:2022hpx, CarrilloGonzalez:2022fwg}).

In particular, the two-sided bounds on all $R^{n-1}c_{n}$ from Ref.~\cite{Heller:2022ejw} tell us already that the landscape lies inside an infinite-dimensional hypercube. Utilising the tools surrounding positive moments, detailed in Ref.~\cite{Bellazzini:2020cot}, we cleave away excluded regions from the hypercube, so as to characterise more precisely the physically relevant region enclosed within. We refer to the region that remains at the end of this cleaving process as {\em the hydrohedron}. Given the minimal set of assumptions that went into this process, the resulting geometry is otherwise completely universal, independent of e.g., spacetime dimension, state, or microscopic theory under consideration.

The hydrohedron has cross-sections of special physical significance. We will consider two examples in detail, corresponding to diffusive and sound modes. We will refer to them as the diffusion cross-section and the sound cross section respectively. In the diffusion cross-section, coefficients of odd powers of $k$ in \eqref{omega_series} are set to zero, while coefficients of even powers are purely imaginary. In the sound cross-section, coefficients of odd powers of $k$ in \eqref{omega_series} are purely real, while coefficients even powers are purely imaginary.

This paper is structured as follows. We begin in Sec. \ref{sec:R} by discussing $R$ and its physical meaning.  Then, in Sec. \ref{sec:diffusion} we present our results for the diffusion cross-section. We analytically construct this cross-section of the full hydrohedron, projected into the space spanned by the three most relevant transport coefficients, leading with diffusivity. Following this we present our results for the sound cross-section in Sec. \ref{sec:sound}, leading with the speed of sound and sound attenuation. In Sec. \ref{sec:otherbounds} we compare our findings to well-known bounds on transport appearing in the literature. We end with a discussion in Sec. \ref{sec:discussion}. The technical aspects of our analysis have been relegated to  appendices.

\section{The natural scale of the problem: $R$}
\label{sec:R}

With the exception of $c_1$, the transport coefficients appearing in \eqref{omega_series} are dimensionful parameters. 
It is conventional to normalise these parameters by thermodynamic quantities, such as appropriate powers of temperature. In the special case of the shear-viscosity $\eta$ (which appears in $\im\,c_2$) the dimensionless combination $\eta/s$ is often considered, where $s$ is the entropy density. The KSS bound $\frac{\eta}{s} \geq \frac{1}{4\pi}$ \cite{Kovtun:2004de} is naturally expressed in this way. 

However, in stark contrast, imposing causality \eqref{cond} gives rise to bounds on transport coefficients normalised by $R$, the radius of convergence of the hydrodynamic series \eqref{omega_series}, as our results below demonstrate.
Roughly speaking, $R$ arises because \eqref{cond} is utilised by integrating within a disk centred on $k=0$, and the strongest constraint is given by the largest disk possible for which the function is still analytic, i.e. a disk of radius $R$. The hydrohedron geometry then lives in the space spanned by the dimensionless transport coefficients, $\{R^{n-1}c_n\}$.

For example, in the case of a shear mode in a conformal theory, $c_2 = -i \frac{\eta}{\epsilon+P}$ we find bounds on the dimensionless combination $R \eta /(\epsilon + P)$. Firstly we find two-sided bounds on $R \eta /(\epsilon + P)$ alone, and then we find an infinite set of bounds which relate $R \eta /(\epsilon + P)$ to other dimensionless transport combinations, $R^{n-1}c_n$. Note that we do not bound the combination $\eta/s$ directly, see also the discussion in Sec. \ref{sec:otherbounds}.

We stress that $R$ is not a formal or abstract quantity. Given a microscopic theory (or a sufficient number of terms of the 
hydrodynamic gradient expansion) it is computable and it is also in principle measurable in experiment. Indeed, $R$ has already been computed in a variety of holographic theories as well as kinetic theory, see \cite{Heller:2020hnq} for a discussion. 
For example, for the ${\cal N} = 4$ SYM at finite temperature and chemical potential $\mu$, a holographic computation gives $R = (\epsilon + P)/(2\mu \sqrt{\eta})$ for a numerically known range of $\mu$ \cite{Withers:2018srf, Jansen:2020hfd}. In general, the value of $R$ varies across theories, spatial dimensions, or within a given theory as the temperature or other thermodynamic parameters are varied. Additionally, $R$ is a natural quantity from an effective field theory point of view; as the radius of convergence of \eqref{omega_series}, it marks the precise point at which nonhydrodynamic degrees of freedom become important. This is because $R$ is set by branch point singularities corresponding to other modes \cite{Heller:2022ejw}. In other words, $R$ is the natural effective field theory cutoff scale for hydrodynamics.
Note that the explicit presence of the UV cutoff in effective field theory bounds should not come as a surprise, see e.g. \cite{Caron-Huot:2020cmc}. 

\section{The diffusion cross-section}
\label{sec:diffusion}

As mention in the Introduction, in this paper we will restrict our analysis to two cross-sections of the full hydrohedron of particular physical significance. We start in this section by specialising to a diffusive mode, that is, a dispersion relation of the form
\be
\omega(k) = i\sum_{n=1}^\infty \beta_{2n} k^{2n}, \label{diffusion_mode}
\ee
where $\beta_{2n} \in \mathbb{R}$ and with a finite radius of convergence, $R>0$. We extend $k$ lie in the disk of radius $R$ centred on $k=0$ in the complex plane, where we impose the causality condition \eqref{cond}. Through a rigorous moment problem analysis we derive a set of hierarchical bounds on the dimensionless coefficients $R^{2n-1}\beta_{2n}$. These bounds define a convex geometry. Full mathematical details of the analysis can found in appendix \ref{app:diffusion}.

The first few orders in this hierarchy of bounds are given by the following expressions
\be
-\frac{16}{3\pi}\leq R\beta_2 \leq 0. \label{diffusion_other}
\ee
\be
-\frac{64}{15\pi} \leq R^3\beta_4  \leq \frac{256 - 15 \pi R\beta_2 (8 + 3 \pi R\beta_2)}{90\pi}.\label{diffusion2}
\ee
\be
\begin{split}
\frac{-32768 + 1575\pi^2\left(R\beta_2 - R^3 \beta_4\right)^2 - 240\pi \left(13 R\beta_2 + 14 R^3 \beta_4\right)}{525\pi\left(16 + 3\pi R \beta_2\right)}\leq R^5 \beta_6 \leq \quad\quad\quad\quad\\
\quad\quad  \frac{4096-525\pi^2 (R\beta_2+ R^3 \beta_4)^2 - 120\pi(31 R \beta_2 + 14 R^3\beta_4)}{175\pi \left(8-3\pi R \beta_2\right)}. 
\end{split} \label{diffusion3}
\ee
Note that only the dimensionless combinations $R^{2n-1}\beta_{2n}$ appear. These bounds are the first three inequalities in an infinite set. Additional inequalities arising at higher orders can be easily derived using the methods outlined in appendix \ref{app:diffusion}, and we restrict providing the first three here for clarity.

\begin{figure}[h!]
    \centering
    \includegraphics[width=0.45\linewidth]{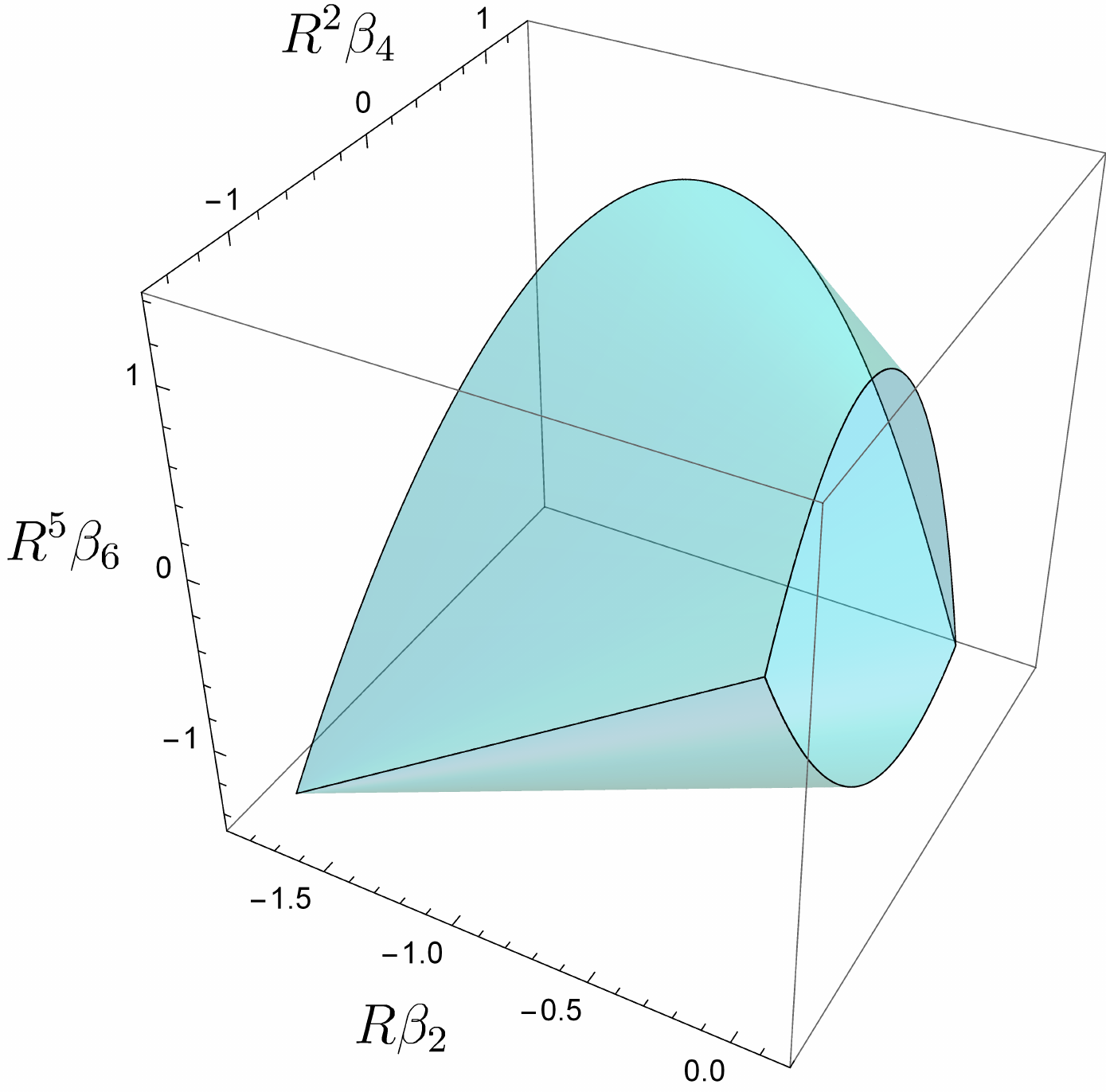}\includegraphics[width=0.45\linewidth]{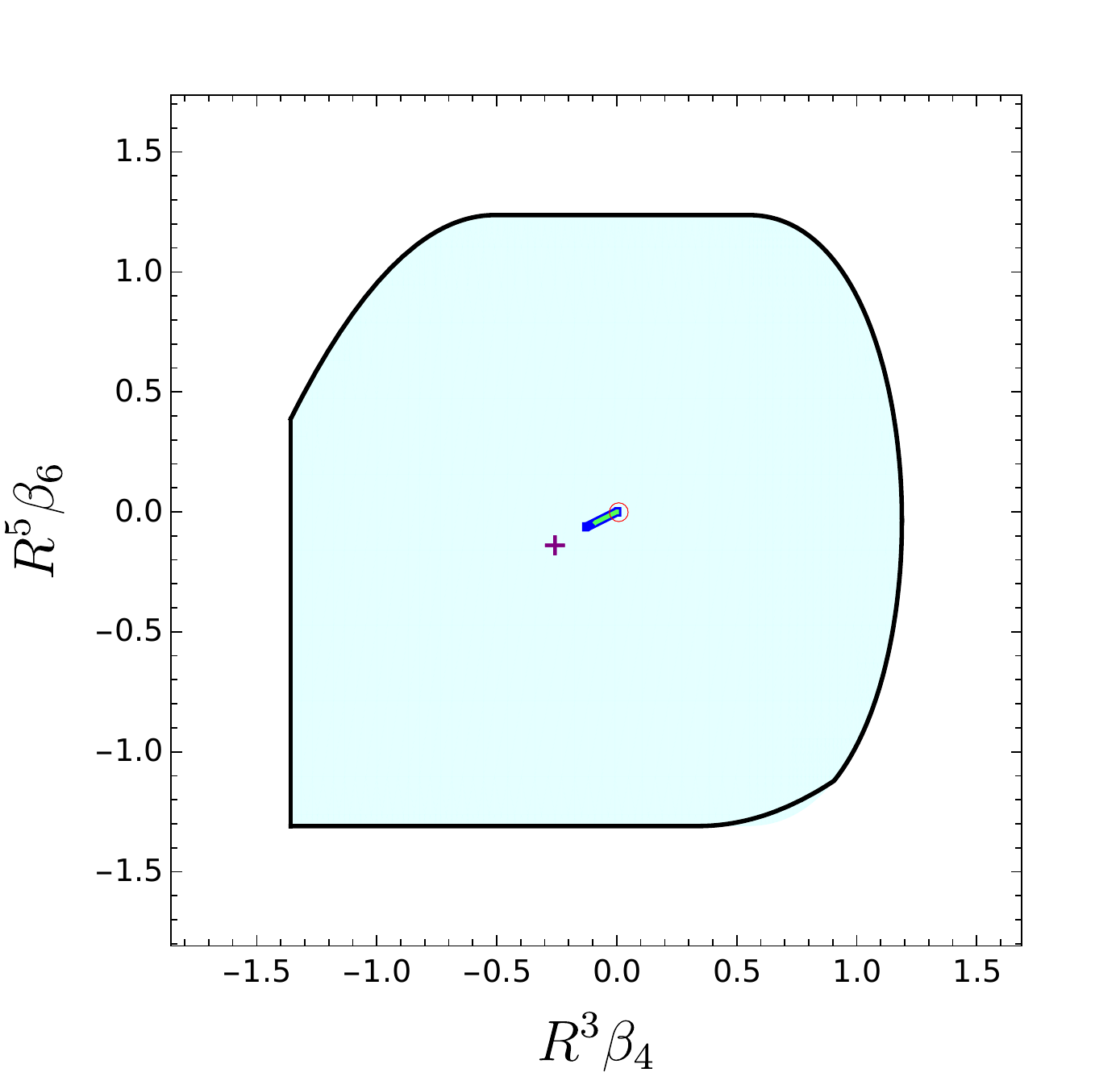}\\
    \includegraphics[width=0.45\linewidth]{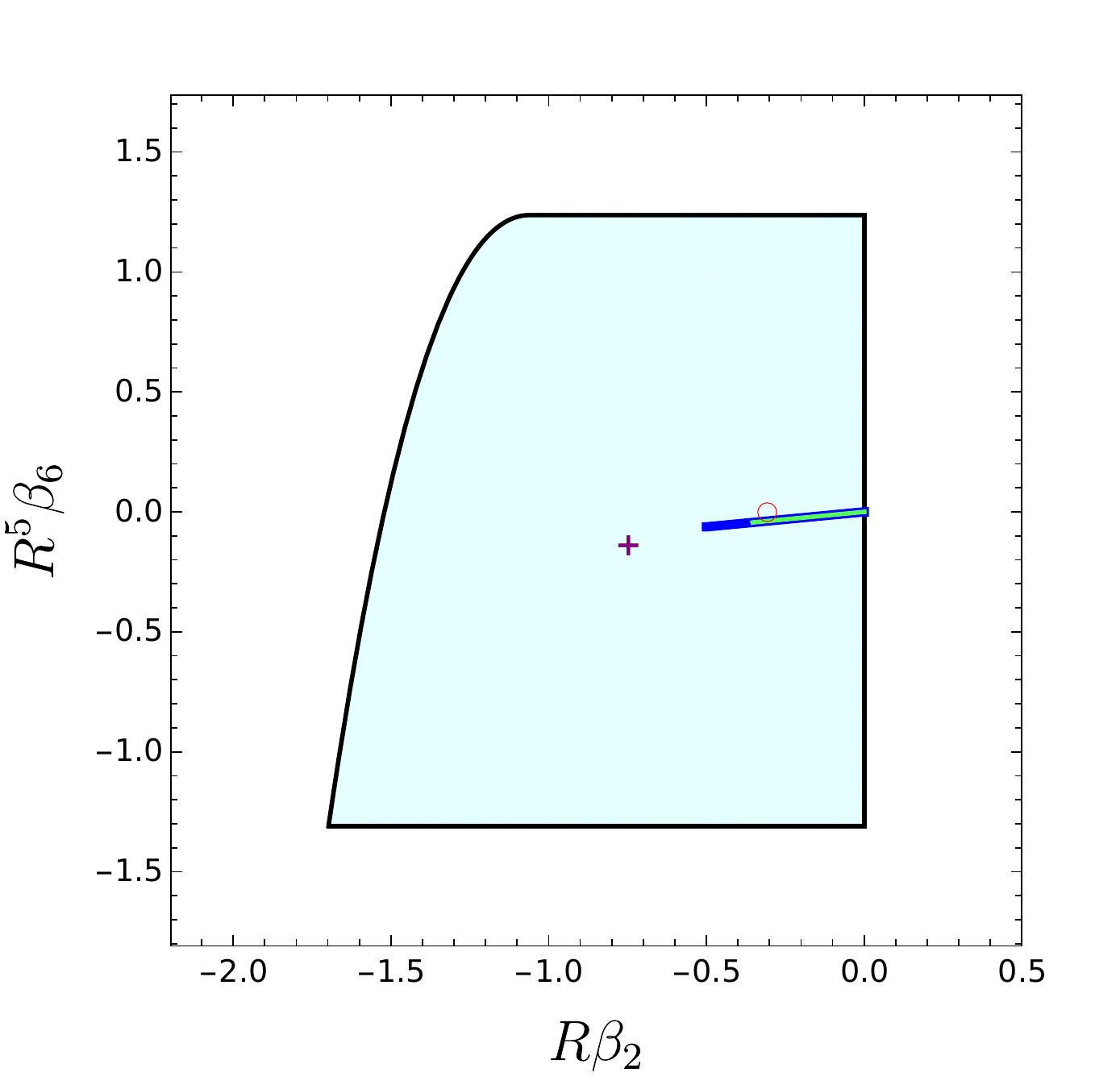}~\includegraphics[width=0.45\linewidth]{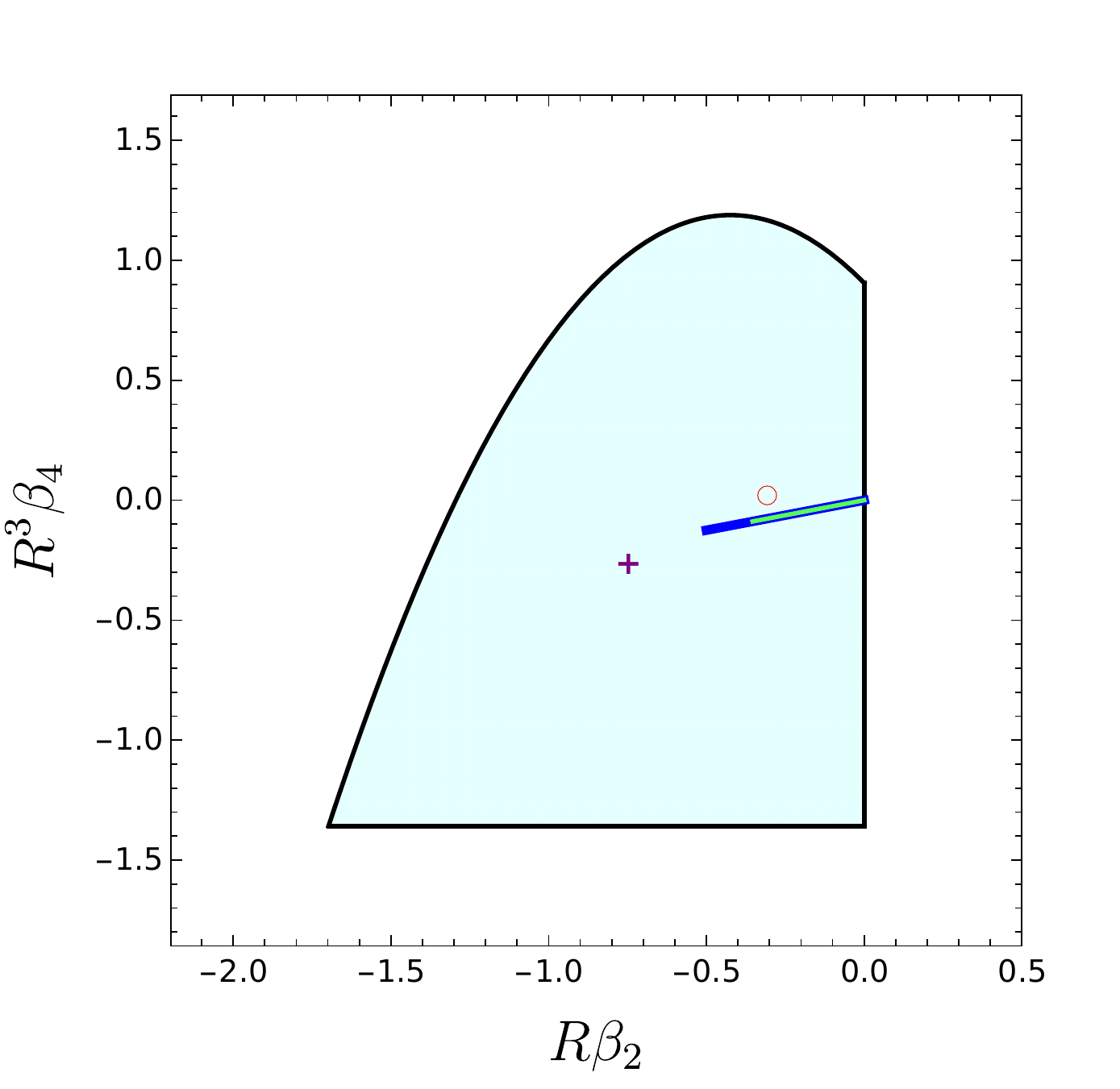}
    \caption{The diffusion mode cross-section of the hydrohedron in the $(R\beta_2, R^3\beta_4, R^5 \beta_6)$-hyperplane is shown in the top-left panel. In the remaining panels we show its projections to the $(R^3\beta_4, R^5\beta_6)$, $(R\beta_2, R^5\beta_6)$ and $(R \beta_2, R^3\beta_4)$-planes (top-right, bottom-left and bottom-right panels respectively). Causal theories necessarily live inside these shaded regions; choices of transport coefficients in the white region are acausal and excluded. The functions determining the boundaries of each projection (solid black lines) are detailed analytically in the main text. The purple crosses represent the location of $\mathcal{N}=4$ SYM in the holographic regime, the red open circles correspond to conformal kinetic theory in the relaxation time approximation, and the green (blue) lines to conformal MIS (conformal BDNK) in the parameter regime where they are causal and linearly stable.}
    \label{diffusionplot}
\end{figure}

Taken altogether, these define a convex geometry in the space of dimensionless transport coefficients $\{R^{2n-1}\beta_{2n}\}$. Since we have restricted our analysis here to diffusive modes \eqref{diffusion_mode}, this is a cross-section of the full hydrohedron geometry (the cross-section corresponding to setting all odd-$k$ coefficients to zero and all even-$k$ coefficients to be purely imaginary). These first three bounds \eqref{diffusion_other} - \eqref{diffusion3} provide a projection of the diffusive cross-section of the hydrohedron to the first three transport coefficients, $\{R\beta_2, R^3\beta_4, R^5\beta_6\}$, a three dimensional convex shape. This is shape is illustrated in figure \ref{diffusionplot} in the top-left panel.

All two-dimensional projections involving $\{R\beta_2, R^3\beta_4, R^5\beta_6\}$ are shown in the other panels of figure \ref{diffusionplot}. The projection to the $(R\beta_2, R^3\beta_4)$-plane is given by the inequalities \eqref{diffusion_other},  \eqref{diffusion2}. The other projection planes $(R\beta_2, R^5\beta_6)$ and $(R^3\beta_4, R^5\beta_6)$ are given by more complicated expressions since they also involve \eqref{diffusion3}. For the $(R\beta_2, R^5\beta_6)$-plane we have that the closure of the projection exists within the interval $-\frac{16}{3\pi} \leq R\beta_2 \leq 0$, with $R^5 \beta_6$ upper bounded by a piecewise function,
\begin{equation}
R^5 \beta_6 \leq \begin{cases}
       \hfil-\frac{3328}{945\pi} - \frac{5}{3}R\beta_2 +\pi R^2\beta_2^2 + \frac{\pi^2}{4} R^3 \beta_2^3 & \text{if} ~ -\frac{16}{3\pi} \leq R\beta_2 < -\frac{10}{3\pi}\\
       \hfil\frac{136}{35\pi} & \text{if} ~ -\frac{10}{3\pi} \leq R\beta_2 \leq 0 \end{cases}, 
\end{equation}
and lower bounded by $-\frac{144}{35\pi}$. For the $(R^3\beta_4, R^5\beta_6)$-plane, the closure of the projection exists within the interval $-\frac{64}{15\pi} \leq R^3 \beta_4 \leq \frac{56}{15\pi}$, with $R^5 \beta_6$ upper bounded by a piecewise function, 
\begin{equation}
R^5\beta_6 \leq  \begin{cases}
        \hfil \frac{512}{175\pi}-\frac{6}{5}R^3 \beta_4 - \frac{3\pi}{8}R^6 \beta_4^2 & \text{if} ~ -\frac{64}{15\pi} \leq R^3 \beta_4 < -\frac{8}{5\pi}\\
        \hfil \frac{136}{35\pi} & \text{if} ~ -\frac{8}{5\pi} \leq R^3 \beta_4 < \frac{26}{15\pi}\\ 
        \hfil\frac{-360+7\sqrt{30}(34+15\pi R^3 \beta_4)\sqrt{56-15\pi R^3\beta_4}}{3150\pi}& \text{if} ~ \frac{26}{15\pi} \leq R^3 \beta_4 \leq \frac{56}{15\pi}
        \end{cases},      
\end{equation}
and lower bounded by another piecewise function,
\begin{equation}
R^5\beta_6 \geq \begin{cases}
         \hfil -\frac{144}{35\pi} & \text{if} ~ -\frac{64}{15\pi} \leq R^3 \beta_4 < \frac{16}{15\pi}\\
         \hfil-\frac{32768 + 105 \pi R^3\beta_4 (32-15\pi R^3\beta_4)}{8400\pi} & \text{if} ~ \frac{16}{15\pi} \leq R^3 \beta_4 \leq \frac{128}{45\pi}\\
         \hfil-\frac{360+7\sqrt{30}(34+15\pi R^3 \beta_4)\sqrt{56-15\pi R^3 \hfil\beta_4}}{3150\pi} & \text{if} ~ \frac{128}{45\pi} \leq R^3 \beta_4 \leq \frac{56}{15\pi}
        \end{cases}. 
\end{equation}

The bounds we outline above, as well as the infinite hierarchy of associated bounds described in appendix \ref{app:diffusion} are a new set of bounds applying to all theories of relativistic transport exhibiting a diffusion mode of the type \eqref{diffusion_mode}. The exception is the upper limit of \eqref{diffusion_other}, which expresses the well-known requirement that the diffusivity $D$ is non-negative, where $D \equiv -\beta_2$. The lower limit of \eqref{diffusion_other}, for instance, is a new rigorous upper bound on diffusion. For any given theory, measuring or computing the $D$ along with the physical microscopic scale $R$, the result necessarily lies inside the bound \eqref{diffusion_other} if the theory is causal.

It is of course instructive to consider where known microscopic theories live in this diffusion cross-section. To this end we show the values of $\{R\beta_2, R^3\beta_4, R^5\beta_6\}$ computed for ${
\cal N} = 4$ SYM theory using holographic techniques,\footnote{See also \cite{Grozdanov:2019kge, Grozdanov:2019uhi} for a discussion of the radius of convergence in ${
\cal N} = 4$ SYM approached using holographic techniques.} conformal kinetic theory in the relaxation time approximation, and two phenomenological models: conformal MIS theory \cite{Muller:1967zza, Israel:1979wp}, and conformal BDNK theory \cite{Bemfica:2017wps, Bemfica:2019knx, Bemfica:2020zjp, Kovtun:2019hdm}. These are generic points in the interior of the projected hydrohedron. The exception is that MIS and BDNK intersect the point where $\beta_{2n} = 0$, where the diffusive mode is trivial. It would of course be interesting to identify nontrivial theories that live at the boundaries of the hydrohedron, where theories live in tension with the constraints of causality. In appendix \ref{app:edges} we demonstrate that many of the boundaries (though not all of them) are open, excluding the possibilities of theories living there.

\section{The sound cross-section}
\label{sec:sound}

In this section we specialize to a mode defined by the following Taylor series expansion of the dispersion relation,
\be
\omega(k) = \sum_{n=0}^\infty \alpha_{2n+1} k^{2n+1} + i\sum_{n=1}^\infty \beta_{2n} k^{2n},  \label{omegasound}
\ee
where $\alpha_{2n+1}$ and $\beta_{2n}$ are real. This includes both sound mode excitations and also Lorentz boosts of the diffusion modes considered in Sec.  \ref{sec:diffusion}. Note that for sound waves $\alpha_1$ is equal to the speed of sound $c_s$, while $\beta_2$ is related to the sound attenuation length $\Gamma_s$ as $\beta_2 = -  \frac{\Gamma_s}{2}$. As before, specialising to modes of the form \eqref{omegasound} will give us a cross-section of the full hydrohedron geometry. Note that diffusion is itself a cross-section of sound along the hyperplane defined by $\alpha_{2n+1} = 0$ for all $n$.

\begin{figure}[h!]
    \begin{center}
    \includegraphics[width=0.45\linewidth]{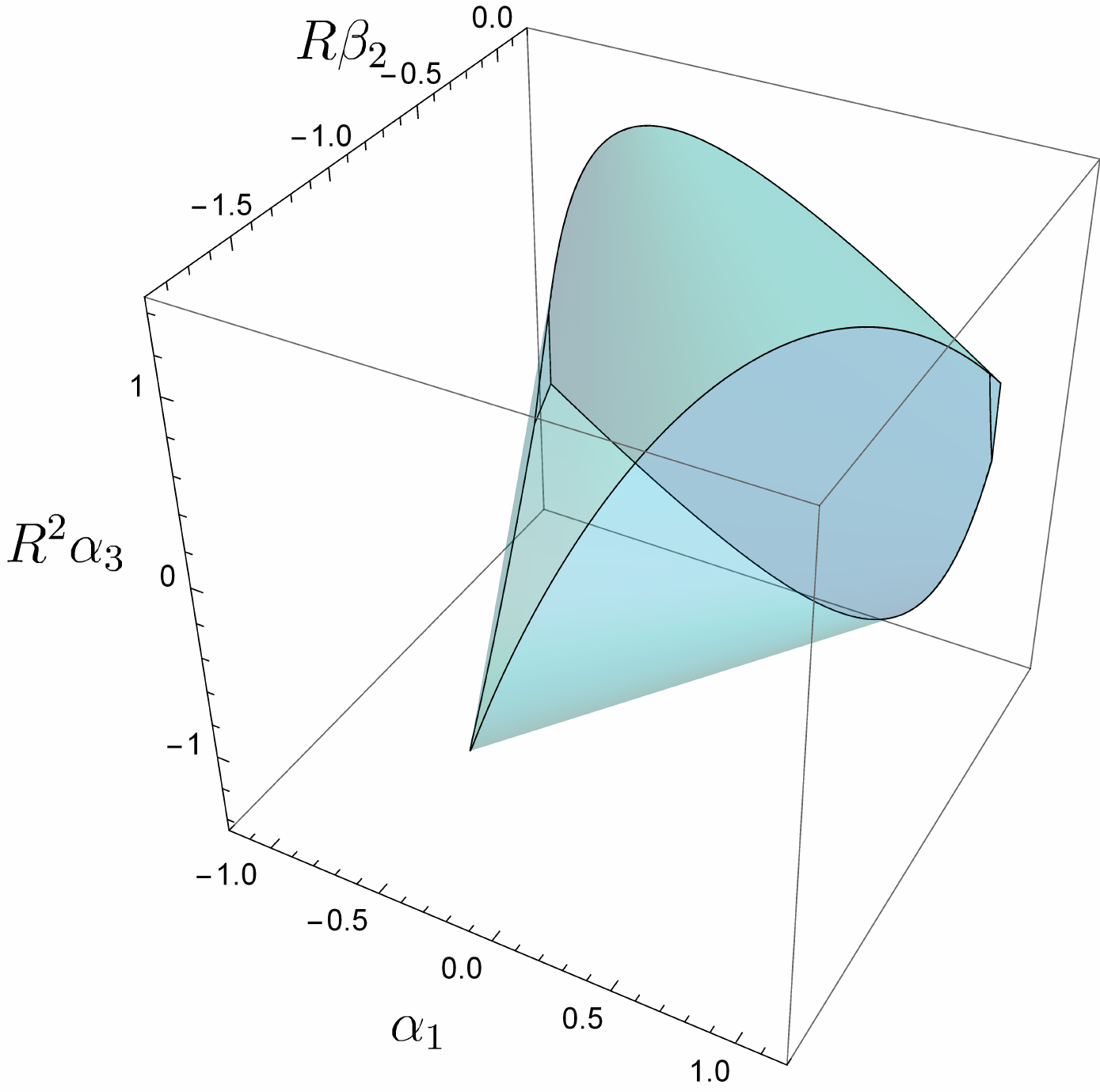}~\includegraphics[width=0.45\linewidth]{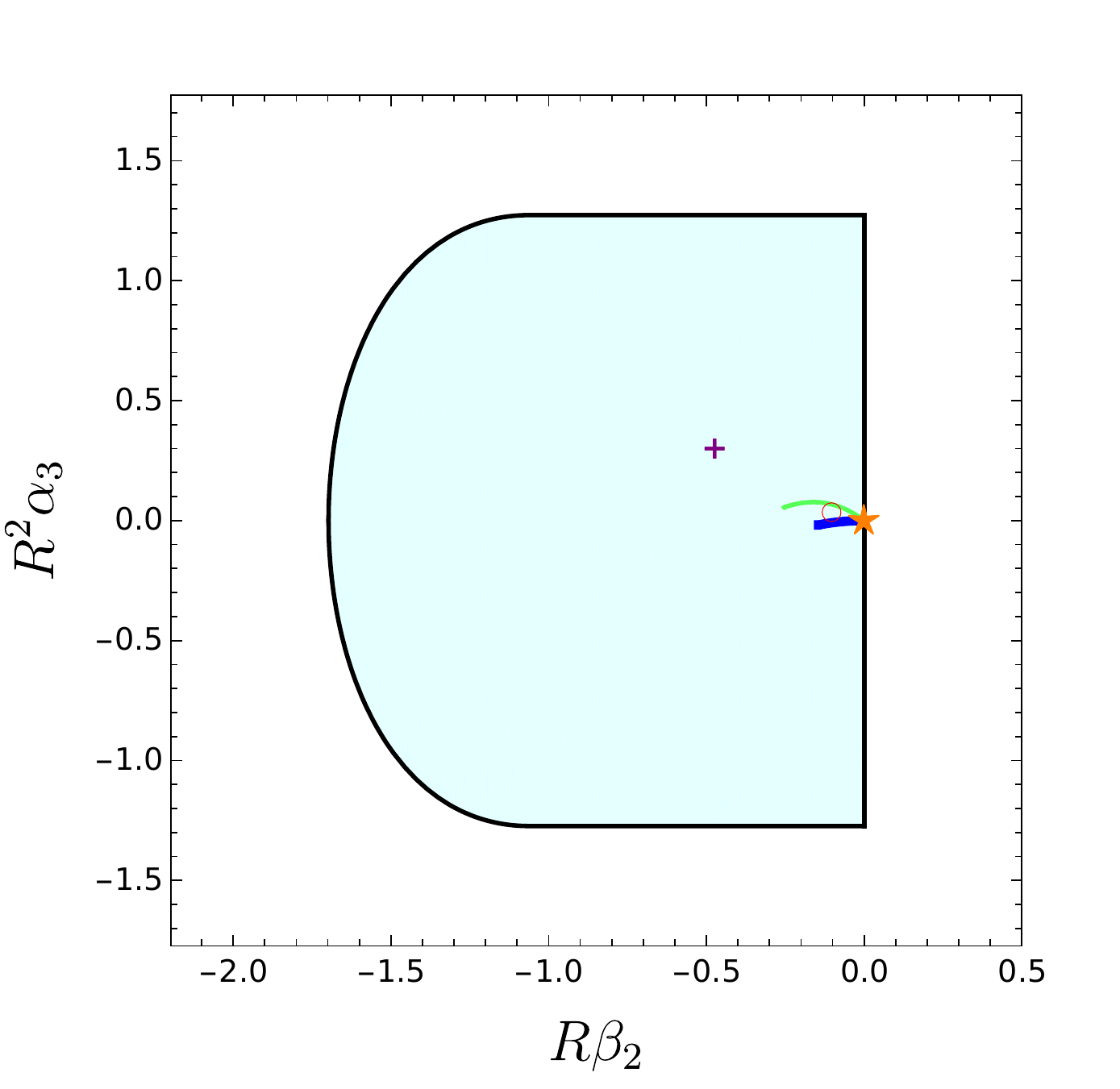}\\
    \includegraphics[width=0.45\linewidth]{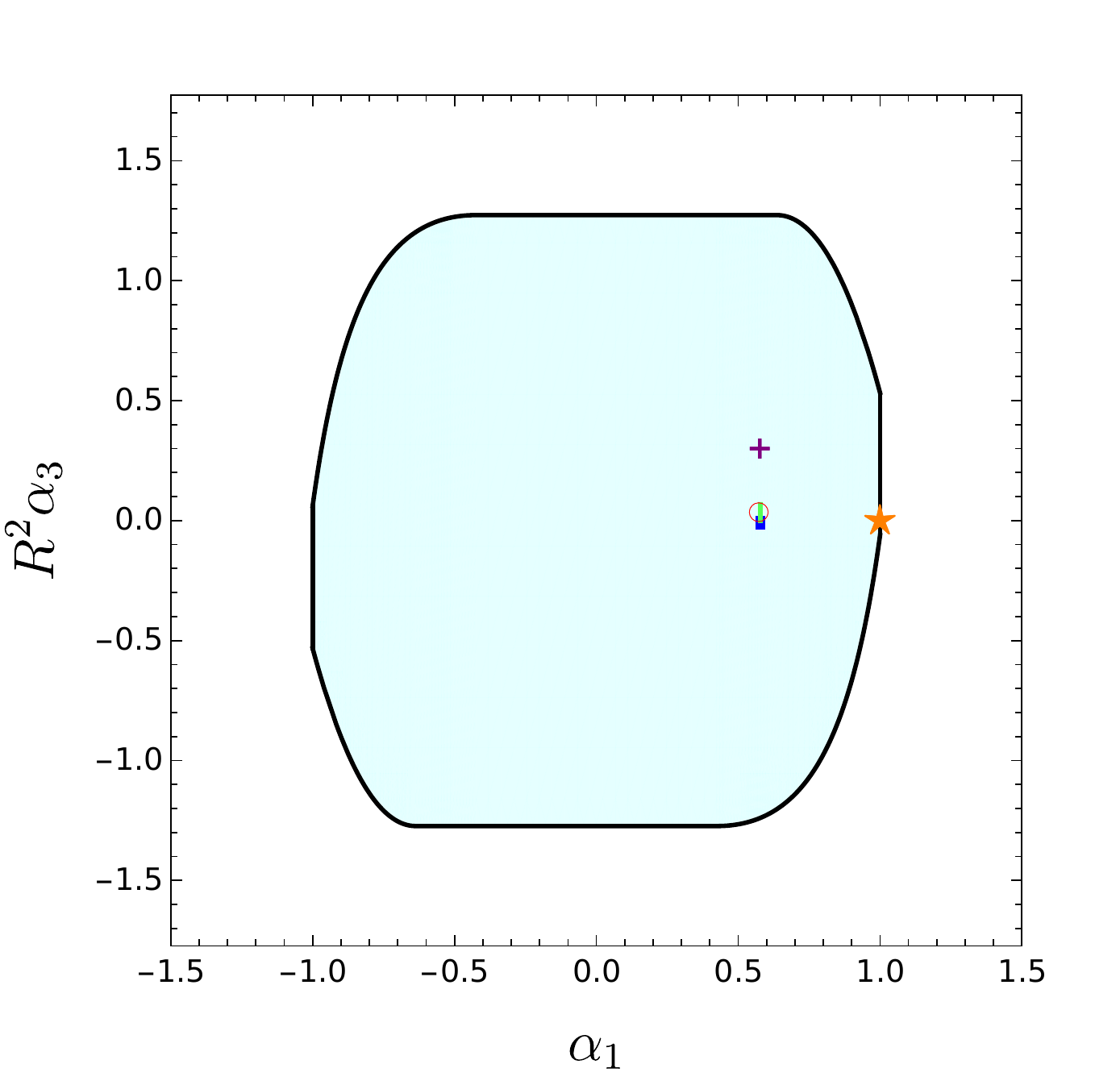}~\includegraphics[width=0.45\linewidth]{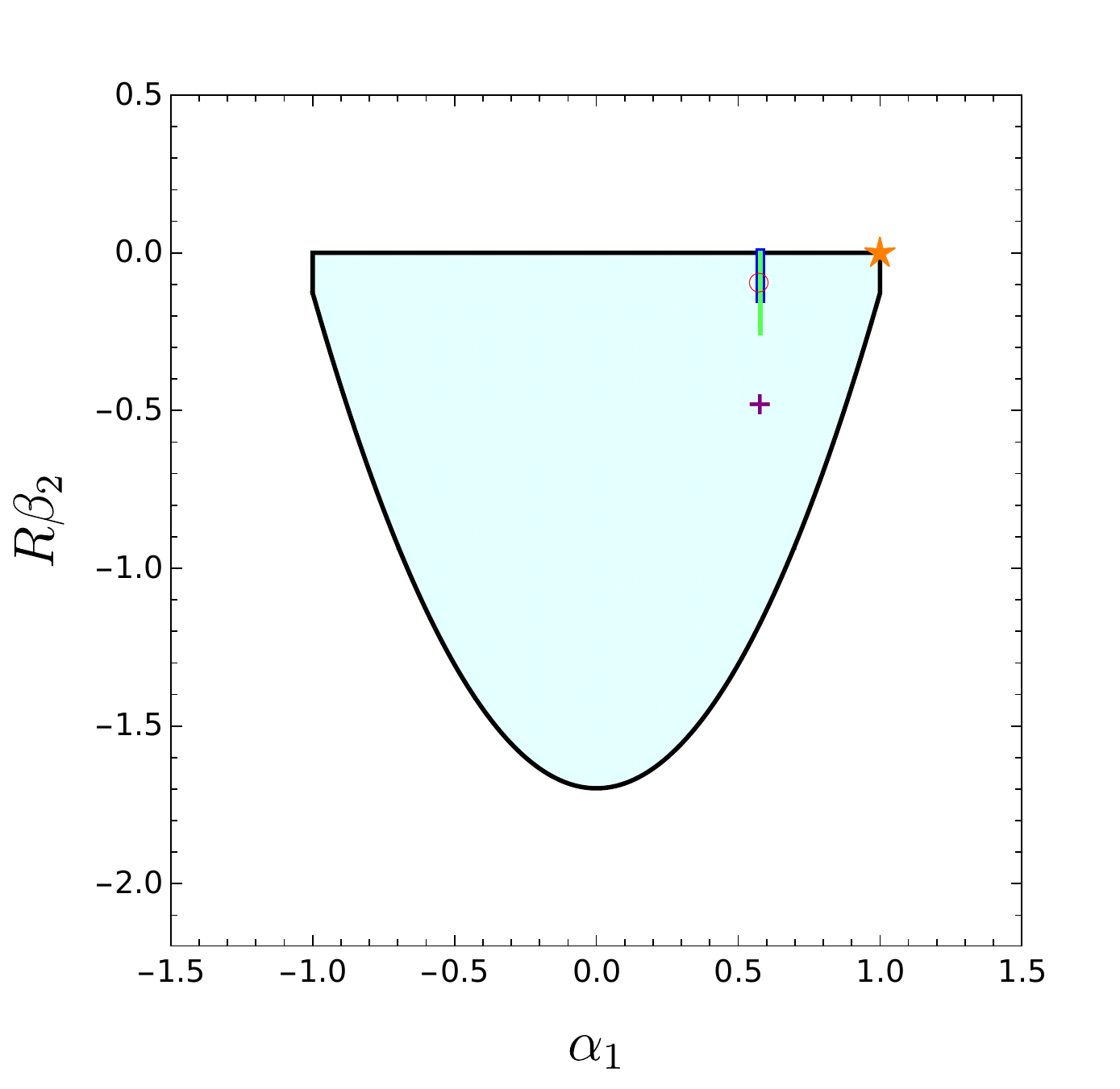}
    \caption{The sound mode cross-section of the hydrohedron in the $(\alpha_1, R \beta_2, R^2 \alpha_3)$-hyperplane (top-left panel), together with its projections to the $(R\beta_2, R^2\alpha_3)$, $(\alpha_1, R^2\alpha_3)$ and $(\alpha_1, R\beta_2)$-planes (top-right, bottom-left and bottom-right panels respectively). Colour coding as in Fig.~\ref{diffusionplot}. The functions determining the boundaries of each projection (solid black lines) are detailed analytically in the main text. The purple crosses represent the location of $\mathcal{N}=4$ SYM in the holographic regime, the red open circles correspond to conformal kinetic theory in the relaxation time approximation, and the green (blue) lines to conformal MIS (conformal BDNK) in the parameter regime where they are causal and linearly stable. Finally, the orange star marks to the special corner point corresponding to the stiff perfect fluid. Note that in these examples for every triplet $(1/\sqrt{3}, \beta_2, \alpha_3)$ there is a corresponding one $(-1/\sqrt{3}, \beta_2, -\alpha_3)$, which we not show explicitly for clarity.}
    \label{soundplot}
    \end{center}
\end{figure}

Through the moment problem analysis outlined in Appendix \ref{app:sound}, we obtain an infinite set of hierarchical bounds on the transport coefficients $\alpha_{2n+1}$, $\beta_{2n}$ normalized to the convergence radius $R$. The first three bounds in this set are:
\begin{equation}\label{a1_bound}
|\alpha_1| \leq 1,     
\end{equation}
\begin{equation}\label{sound2}
-\frac{16}{3\pi} + \frac{\pi}{2}\alpha_1^2 \leq R \beta_2 \leq  0,   \end{equation}
\begin{equation}\label{sound3}
\begin{split}
\frac{128-9\pi^2(\alpha_1-R\beta_2)^2 - 12\pi(\alpha_1+2R\beta_2)}{9\pi(-4+\pi\alpha_1)} \leq R^2\alpha_3 \leq
\frac{128 - 9\pi^2(\alpha_1+R\beta_2)^2 + 12 \pi(\alpha_1-2R\beta_2)}{9\pi(4+\pi\alpha_1)}.
\end{split}
\end{equation}
The first bound \eqref{a1_bound} expresses the well-known fact that in a causal theory the speed of sound cannot exceed the lightcone speed $v_{\text{LC}} = 1$. The remaining two-sided bounds \eqref{sound2} and \eqref{sound3} are new. In particular, if the speed of sound is known, the inequality \eqref{sound2} provides an upper bound for the sound attenuation length $\Gamma_s$ in units of the convergence radius. For instance, in a $d$-dimensional conformal field theory, $|\alpha_1|=\frac{1}{\sqrt{d-1}}$ and
\begin{equation}
R\Gamma_s \leq \frac{32}{3\pi} - \frac{\pi}{d-1}. 
\end{equation}
Taken together, the bounds \eqref{a1_bound}-\eqref{sound3} define a projection of the sound cross-section of the full hydrohedron geometry to the three-dimensional subspace of the first three transport coefficients $\{\alpha_1, R\beta_2, R^2\alpha_3\}$. This projection is a convex shape illustrated in the top-left panel of Fig.~\ref{soundplot}.

All two-dimensional projections involving pairs of $\{\alpha_1, R\beta_2, R^2\alpha_3\}$ are shown in the remaining panels of Fig.~\ref{soundplot}. The projection to the $(\alpha_1,R\beta_2)$-plane is given by inequalities \eqref{a1_bound} and \eqref{sound2}. As it happened in the diffusion case, the projections to the remaining $(\alpha_1,R^2\alpha_3)$ and $(R\beta_2,R^2\alpha_3)$ planes involve a higher-level inequality, \eqref{sound3}, and have therefore more involved explicit expressions. In the $(\alpha_1,R^2\alpha_3)$-plane, we find that the closure of the projection exists within the interval $|\alpha_1|\leq 1$, in which $R^2 \alpha_3$ is upper bounded by a piecewise function, 
\begin{equation}
R^2 \alpha_3 \leq \begin{cases}
      \hfil \frac{128+3\alpha_1\pi (4-3\pi \alpha_1)}{9\pi(4+\pi\alpha_1)} & \text{if} ~ -1 \leq \alpha_1 < -\frac{4}{3\pi}\\
      \hfil \frac{4}{\pi} & \text{if} ~ -\frac{4}{3\pi} \leq  \alpha_1 < \frac{2}{\pi} \\
      \hfil \alpha_1 \left(3-\frac{\pi^2}{4}\alpha_1^2\right) & \text{if} ~ \frac{2}{\pi} \leq \alpha_1 \leq 1
    \end{cases},     
\end{equation}
and lower bounded by another piecewise function,
\begin{equation}
R^2 \alpha_3 \geq \begin{cases}
      \hfil \alpha_1 \left(3-\frac{\pi^2}{4}\alpha_1^2\right) & \text{if} ~ -1\leq\alpha_1< -\frac{2}{\pi} \\
      \hfil -\frac{4}{\pi} & \text{if} ~ -\frac{2}{\pi}\leq\alpha_1<\frac{4}{3\pi} \\
      \hfil -\frac{128-3\alpha_1\pi (4+3\pi \alpha_1)}{9\pi(4-\pi\alpha_1)} & \text{if} ~ \frac{4}{3\pi} \leq \alpha_1 \leq 1
    \end{cases}.  
\end{equation}
For the $(R \beta_2, R^2 \alpha_3)$-plane, the closure of the projection exists within the interval  $-\frac{16}{3\pi} \leq R\beta_2 \leq 0$, where $R^2\alpha_3$ is upper bounded by a piecewise function, 
\begin{equation}
R^2\alpha_3\leq \begin{cases}
      \hfil \frac{(2-3\pi R\beta_2)(96+18\pi R\beta_2)^\frac{1}{2}}{18\pi} & \text{if} ~ -\frac{16}{3\pi} \leq R\beta_2 < -\frac{10}{3\pi}\\
      \hfil \frac{4}{\pi} & \text{if} ~ -\frac{10}{3\pi}\leq R\beta_2 \leq 0
    \end{cases},      
\end{equation}
and lower bounded by another piecewise function, 
\begin{equation}
R^2\alpha_3\geq \begin{cases}
      \hfil -\frac{(2-3\pi R\beta_2)(96+18\pi R\beta_2)^\frac{1}{2}}{18\pi} & \text{if} ~ -\frac{16}{3\pi} \leq R\beta_2 < -\frac{10}{3\pi}\\
      \hfil -\frac{4}{\pi} & \text{if} ~ -\frac{10}{3\pi}\leq R\beta_2 \leq 0
    \end{cases}. 
\end{equation}

In parallel with diffusion case, in Fig.~\ref{soundplot} we also provide the values of $\{\alpha_1, R\beta_2, R^2\alpha_3 \}$ for $\mathcal{N}=4$ SYM in the holographic limit, kinetic theory in the relaxation time approximation, MIS and BDNK (the last three also in the conformal regime). Note that $\alpha_1 = \frac{1}{\sqrt{3}}$ in all cases. Again, we find that all of these theories lie at generic points inside the hydrohedron projections, the exception being where MIS and BDNK reach the boundary at $\beta_{2n} = 0$ where the sound mode is that of a perfect conformal fluid.

Finally, we note that the sound cross-section has a special corner point. As we demonstrate in Appendix \ref{app:perfectfluid},  when $|\alpha_1|=1$ all the higher-order transport coefficients vanish and the dispersion relation is uniquely determined, 
\be
\omega(k) = \pm k. \label{stiffeq}
\ee
This corresponds to the case of a stiff perfect fluid, and fluids boosted to the speed of light, since \eqref{stiffeq} is a fixed point under boosts.\footnote{The vanishing of the diffusivity for a stiff fluid was also noted recently in Ref.~\cite{Gavassino:2023myj}. 
Note also that, according to our analysis, $\alpha_1 = \pm 1$ still allows for a range of $R\beta_2$ and $R^2\alpha_3$ values. It is possible that such points may be understood in a limit of a family of dispersion relations where as $\alpha_1 \to \pm 1$ one finds $R\to \infty$ and $\beta_2, \alpha_3\to 0$ but with the dimensionless products $R\beta_2$ and $R^2\alpha_3$ finite in this limit.} Additionally, there are various other physical theories that live at this corner of the hydrohedron, traditionally outside the realm of hydrodynamics, for which our analysis still applies. These include free massless theories and 2d CFTs.

\section{Relation to other bounds} \label{sec:otherbounds}

We have obtained an infinite new class of bounds on transport coefficients. In the literature there are a number of important bounds on transport that have appeared before. In this section we comment on the relation to these other bounds.

Perhaps the closest in spirit is that obtained in \cite{Hartman:2017hhp} for a qualitative upper bound on diffusion following from causality considerations,
\be
D \lesssim v^2 \tau_\text{eq}, \label{HHM}
\ee
where $v$ is a speed defining a lightcone for operator growth and $\tau_\text{eq}$ is a local equilibration time.
This bound was obtained using approximations for the structure of diffusion and gapped modes. This upper bound should be compared with the lower bound in \eqref{diffusion_other} in the present work, i.e.
\be
D \leq \frac{16}{3\pi} v_{\text{LC}} R^{-1} \label{ourHHM}
\ee
which can be viewed as a rigorous version of \eqref{HHM}. Recall here that $v_\text{LC}$ is the lightcone speed. Note that the rigorous version demands that equilibration time is replaced by radius of convergence through $v^2 \tau_\text{eq} \rightarrow v_\text{LC} R^{-1}$, and the precise coefficient is determined in \eqref{ourHHM}. As we discuss in Sec. \ref{sec:R}, $R$ and $\tau_\text{eq}^{-1}$ are both characterisations of the scale of non-hydrodynamic physics, and so it is natural that this replacement appears here.

Another notable bound on transport is the KSS bound \cite{Kovtun:2004de}, a conjectured lower bound on viscosity in units of entropy density,
\be
\label{eq.KSS}
\frac{1}{4\pi} \leq \frac{\eta}{s}.
\ee
We know from string theory considerations \cite{Buchel:2008vz} that the value of $\eta/s$ can be lowered below this value, at least perturbatively.
We have no direct analogue for this bound. This illustrates that the KSS bound (or its improved version taking the results of~\cite{Buchel:2008vz} into account) does not follow from causality alone\footnote{This is in line with the speed of light $v_{LC}$ not appearing at all in~\eqref{eq.KSS}.}, but, if at all, from other considerations. The closest would be the upper bound in \eqref{diffusion_other}, in which we conclude that $0 \leq D R$.

Finally we comment on relations to other Planckian bounds related to KSS, as proposed in \cite{Hartnoll:2014lpa, Blake:2016wvh, Blake:2016sud, Hartnoll:2021ydi}. These are again lower bounds on diffusion, which involve the butterfly velocity $v_B$ and the Lyapunov time $\tau_L$, for instance
\be
v_B^2 \tau_L\lesssim D.
\ee
Similar comments apply to these bounds; as lower bounds that do not appear in our analysis it seems unlikely that they are a consequence of the constraints of causality.

\section{Discussion} \label{sec:discussion}
In this paper we provide a new look at relativistic hydrodynamics, focusing on constraining the theory as much as possible based on fundamental principles alone. Using microscopic causality we exclude most of transport coefficient space, leaving a convex geometry in which all causal theories of hydrodynamics necessarily reside. This geometry is uniquely determined from causality alone and thus universal, independent of spacetime dimension, state, and any microscopic details. We provide constraints at all orders in the hydrodynamic expansion, and present a detailed analysis of the first few orders in the sound and diffusion cross-sections of this geometry.

It is important to note that the hydrohedron is a geometry defined in the space of transport coefficients in units of the momentum scale $R$, where $R$ is the radius of convergence of the hydrodynamic Taylor series \eqref{omega_series}. This is an intrinsic scale that may be computed or measured given a specific theory. This may appear an unfamiliar normalisation choice, given that the vast majority of previous hydrodynamic literature quotes transport coefficients in units such as temperature, or entropy density. However, the {\em universal} hydrohedron geometry arises only in units of $R$. Since in general $R$ is a function of temperature in a way that depends on the theory, if one were to convert units then the shape would be different between theories and universality would be lost. Indeed, the scale $R$ is the cutoff for the effective theory, marking the breakdown of hydrodynamics where other physical degrees of freedom are required. It is only natural therefore that the universal hydrohedron geometry is apparent when coefficients are normalised by $R$.

Our analysis did not depend on being in the fluid rest frame. While the diffusion mode analysis of Sec. \ref{sec:diffusion} was specialised to zero background fluid velocity, applying a boost $v$ to the dispersion relation \eqref{diffusion_mode} turns it into the form of a sound mode dispersion relation \eqref{omegasound}. Transport coefficients can be converted straightforwardly; for first few orders, $\alpha_1 \rightarrow v$, $\beta_2 \rightarrow (1-v^2)^{\frac{3}{2}}\beta_2$, $\alpha_3 \rightarrow 2 v(1-v^2)^2\beta_2^2$. The radius of convergence $R$ does not boost in a easily predictable way and requires a microscopic computation in each case. The sound mode results of Sec. \ref{sec:sound} then apply.

We saw that some faces of the hydrohedron were excluded, meaning that the set is open there. Examples are the faces which lie at the boundary of the regions excluded by the moment problem. This is not the case for all faces however, with the faces $\alpha_1 = \pm 1$ and $\beta_2 = 0$ not following from the moment problem. This raises the interesting possibility of `distinguished' theories living there, reminiscent of the 3D Ising model in the conformal bootstrap~\cite{El-Showk:2012cjh}. Indeed, we proved that on the faces $\alpha_1 = \pm 1$ the dispersion relation is uniquely $\omega(k) = \pm k$, i.e. a stiff perfect fluid. For the face $\beta_2 = 0$ we have found examples of theories which live there\footnote{A family of examples is given by $\omega(k) = 4(1+i)\beta_4\left(i - 1 + \sqrt{1+i k^2} - i \sqrt{1+k^2}\right)$, for $-1/(4\sqrt{2})\leq \beta_4 \leq 0$.}, but none that appear to be uniquely determined by low order transport coefficients (aside from at the intersection with $\alpha_1 = \pm 1$). It would be interesting to see if there are other geometrically privileged points corresponding to theories of special significance.

Our results use analyticity in two senses. Firstly, the role of \eqref{cond} is simply to exclude singularities from the analytic domain of momentum space Green's functions, which follow directly from the axioms of quantum field theory, as discussed in \cite{Heller:2022ejw}. Secondly, we required $\omega(k)$ be analytic at $k=0$ \eqref{omega_series}, i.e. the classical hydrodynamic expansion. However, more is known about the analytic structure of $\omega(k)$ from first principles, for instance it cannot contain poles \cite{Heller:2022ejw}. Thus, finding a way to incorporate properties of the `global' analytic structure of $\omega(k)$ -- rather than just analyticity at $k=0$ -- may result in a more constrained region of transport coefficient space. For instance, a recent conjecture for the analytic structure of chaotic large $N$ thermal two-point functions~\cite{Dodelson:2023vrw} may have bearing on the analytic structure of $\omega(k)$.
Analogously, it is a conjecture in the S-matrix bootstrap programme that the amplitudes are `maximally analytic', and given this additional information allows for stronger constraints; see for example \cite{Paulos:2017fhb, Correia:2020xtr, Kruczenski:2022lot}. 

This discussion brings us naturally to the problem of including and understanding stochastic fluctuations in this language. In this case it is known that $\omega(k)$ becomes non-analytic at $k=0$, see for example \cite{Kovtun:2012rj}.
These effects arise from nonlinearities treated in perturbation theory, for example, correcting $\omega \to \omega + g\, \delta \omega_{\text{stochastic}} + O(g)^2$ where $g$ is a coupling constant which controls interaction strength. Therefore, as long as the perturbative treatment holds, such corrections can only become important when the causality inequality \eqref{cond} is saturated, which then imposes that $\im(\delta \omega_{\text{stochastic}}) \leq 0$. It would also be interesting to investigate stochastic effects at the level of the Schwinger-Keldysh effective action for hydrodynamics \cite{Haehl:2015foa, Crossley:2015evo, Jensen:2017kzi, Haehl:2018lcu, Liu:2018kfw}. In this context it seems natural to attempt to lift the techniques presented here to the coefficients that appear in this effective action instead. 

Finally, the methods adopted in our paper can be also used to chart causal convex geometries for other mode types such as Goldstone modes, fast decaying excitations -- such as transient quasinormal modes of holographic black holes -- and quasiparticles.

\section*{Acknowledgements}
It is a pleasure to thank F. Haehl and S. Hartnoll for discussions. 
A.\,S. acknowledges financial support from Grant No. CEX2019-000918-M funded by Ministerio de Ciencia e Innovación (MCIN)/Agencia Estatal de Investigación (AEI)/10.13039/501100011033. M.\,S. is supported by the National Science Centre, Poland, under Grant No. 2021/41/B/ST2/02909. B.\,W. is supported by a Royal Society University Research Fellowship and in part by the Science and Technology Facilities Council (Consolidated Grant `Exploring the Limits of the Standard Model and Beyond'). This project has received funding from the European Research Council (ERC) under the European Union’s Horizon 2020 research and innovation programme (grant number: 101089093 / project acronym: High-TheQ). Views and opinions expressed are however those of the authors only and do not necessarily reflect those of the European Union or the European Research Council. Neither the European Union nor the granting authority can be held responsible for them. 

\appendix

\section{The moment problem}

\subsection{Diffusion cross-section derivation}\label{app:diffusion}

A general diffusive mode can be written as the Taylor series \eqref{diffusion_mode}. Inserting \eqref{diffusion_mode} into Eq.~\eqref{cond} and taking the $r\to 0$ limit at $\theta = 0$ it immediately follows \cite{Heller:2022ejw} that the diffusion constant must be non-negative\footnote{If $\beta_2 = 0$ then one can also show $\beta_{4} \leq 0$, and so on in this fashion.}
\be
\beta_2 \leq 0. \label{diffusion_trivial}
\ee
To obtain additional bounds on transport we may multiply \eqref{cond} by any non-negative periodic function of $\theta$, $\tilde{p}(\theta)$ and integrate around a circle of radius $r<R$,
\be
\int_{0}^{2\pi} \tilde{p}(\theta)\tilde{\mu}(\theta) d\theta\geq 0, \qquad \tilde{\mu}(\theta) \equiv \frac{|\im\, k|-\im\,\omega(k)}{|k|},
\ee
or, more conveniently for $x = \cos\theta$,  
\be
\int_{-1}^{1} p(x)\mu(x) dx\geq 0, \quad \mu(x) \equiv  1 - \sum_{n=1}^\infty r^{2n-1}\beta_{2n} (1-x^2)^{-\frac{1}{2}}T_{2n}(x), \label{diffusion_measure}
\ee
where $T_n(x)$ are Chebyshev polynomials of the first kind.
Given some $p(x)$, the left-hand side of \eqref{diffusion_measure} evaluates to a sum of transport coefficients, revealing a new bound. Assembling all such bounds, one can then carve out regions of excluded parameter space by considering all $r<R$. 

An analogous construction in the context of scattering amplitudes is discussed in Ref.~\cite{Bellazzini:2020cot} utilising the theory of moments, an efficient route to generate optimal bounds, which we adopt here. Since $\mu(x) = \mu(-x)$, we restrict our attention to even functions of $x$.
Then, the condition \eqref{diffusion_measure} becomes a condition on the matrices of moments of the measure $\mu(x) dx$. In particular, at a given $N$, we construct the following Hankel matrices of the moments of this measure \cite{Bellazzini:2020cot,schmudgen2020lectures},
\be
\left(H^\ell_N\right)_{ij} = a_{i+j+\ell},\quad i,j = 0,\ldots,\biggr\lfloor \frac{N-\ell}{2} \biggr\rfloor,
\qquad a_n \equiv \frac{1}{2}\int_{-1}^1 x^{2n} \mu(x) dx,
\ee
then \eqref{diffusion_measure} is the condition that the following matrices are positive semi-definite,
\be
H^0_N \succeq 0, \quad H^1_N\succeq 0, \quad H^0_{N-1} - H^1_N\succeq 0, \quad H^1_{N-1} - H^2_{N} \succeq 0.
\ee
The moments $a_n$ are related to the transport coefficients $\beta_{2n}$ \eqref{diffusion_measure} through
\bea
a_n &=& \frac{1}{2n+1} - 2^{-(2n+1)} \pi \sum_{j=1}^n  \binom{2n}{n-j} r^{2j-1}\beta_{2j},\label{moment_formula}  \\
r^{2n-1}\beta_{2n} &=& \frac{4}{\pi (1-4 n^2)} - \frac{4}{\pi} \sum_{j=0}^n\sum_{q=0}^{j} (-1)^{j-q} \binom{2n}{2j} \binom{j}{q}a_{n-j+q},\label{inversion_formula}
\eea
and thus given $N$ the conditions of positive semi-definite Hankel matrices translate into bounds on a subset of transport coefficients. 
Here we consider the constraints arising from the $N=2$ and $N=3$ moment problems, and thus construct the projections of the hydrohedron into parameter space spanned by the transport coefficients $\{R\beta_2, R^3\beta_4, R^5\beta_6\}$ described in the main text. 

\paragraph{\underline{$N=2$ case:}} In this case we require positive semi-definiteness of the following matrices for all $r<R$,
\be
\begin{gathered}
H^0_2 = \begin{pmatrix}
1 & \frac{1}{3} - \frac{\pi}{8}r\beta_2\\
\frac{1}{3} - \frac{\pi}{8}r\beta_2 & \frac{1}{5} -\frac{\pi}{8}r\beta_2 - \frac{\pi}{32}r^3\beta_4     
\end{pmatrix}, \\
 H^1_2 = \begin{pmatrix}
\frac{1}{3} - \frac{\pi}{8}r\beta_2 
\end{pmatrix},\quad
H^0_1 - H^1_2 = \begin{pmatrix}
\frac{2}{3} + \frac{\pi}{8}r\beta_2 
\end{pmatrix}, \quad
 H^1_1 - H^2_2 = \begin{pmatrix}
\frac{2}{15} + \frac{\pi}{32}r^3\beta_4 
\end{pmatrix}. 
\end{gathered}
\ee
These conditions, together with inequality \eqref{diffusion_trivial},  lead to the inequalities \eqref{diffusion_other} and \eqref{diffusion2}. 

It is straightforward to demonstrate that the non-excluded region defined by inequalities \eqref{diffusion_other} and \eqref{diffusion2} is not closed. For example, at $\beta_2 = 0$ we have that $\beta_4 \leq 0$, and thus a portion of the $\beta_2 = 0$ boundary indicated in the bottom-right panel of Fig.~\ref{diffusionplot} is excluded. With greater effort it can be established that the quadratic portion of this boundary as well as the line $\beta_4 = -\frac{64}{15\pi}$ are also excluded. At these loci in the $(R\beta_2, R^3\beta_4)$-plane, by considering the constraint \eqref{cond} at $r=R$ uniquely fixes all higher-order transport coefficients. This leads to dispersion relations which feature poles and hence violate the causality condition \eqref{cond} (see Ref.~\cite{Heller:2022ejw}), thus showing that loci where the first inequality in \eqref{diffusion2} is saturated do not belong to the hydrohedron. An example of such dispersion relation is 
\begin{align}
\ww(\kk) &= - \frac{2i}{\pi} \left((\kk-\kk^{-1})\arctanh{\kk} + \frac{1-2(1-2a_1)\kk^2 + \kk^4}{1-\kk^4}\right) &&\text{($\beta_4 = -64/(15\pi)$)}\label{diffusion_linear_boundary}
\end{align}
with $k = R \kk$, $\omega = R \ww$, and $a_1 = \frac{1}{3}-\frac{\pi}{8}\beta_2 \in [1/3,1]$ parameterising the curve in the projection plane.  Further details may be found in Appendix \ref{app:edges}. 

\paragraph{\underline{$N=3$ case:}} $H^0_3\succeq 0$ and $H^1_2-H^2_3\succeq 0$ give no new constraints beyond those in \eqref{diffusion2}, whereas $H^0_2-H^1_3\succeq 0$ and $H^1_3\succeq 0$ respectively, for all $r<R$, give the additional bounds \eqref{diffusion3}. 

We note that when $R^5\beta_6$ saturates any of the inequalities in Eq.~\eqref{diffusion3}, both $R^5\beta_6$ and all the higher-order transport coefficients are fixed in terms of $R\beta_2$ and $R^3 \beta_4$. The resulting dispersion relations are excluded due to the presence of poles (see Appendix \ref{app:edges}). As a consequence, the loci where Eq.~\eqref{diffusion3} is saturated do not belong to the hydrohedron.

\subsection{Sound cross-section derivation}\label{app:sound}

The dispersion relation of a sound mode has the Taylor series representation \eqref{omegasound}. The bound on the speed of sound \eqref{a1_bound} is obtained by considering the fundamental inequality \eqref{cond} at $\theta=\pi/2$ in the $r \to 0$ limit \cite{Heller:2022ejw}. As it happened in the diffusion case, to find the remaining ones the optimal way to proceed is translating the causality condition \eqref{cond} into a moment problem. This time, however, the relevant moment problem is a trigonometric one formulated on the circle $\theta \in [0, 2\pi)$. We work with the following unit-normalized density
\begin{equation}
\mu(\theta) \equiv \frac{|\im\, k|-\im\,\omega(k)}{4|k|}\biggr\rvert_{k = r e^{i\theta}},      
\end{equation}
and consider the moments 
\begin{equation}\label{gamma_def}
\gamma_n \equiv \int_0^{2\pi} e^{-i n \theta} \mu(\theta)d\theta, \quad n=0,\pm 1, \pm 2, \ldots    
\end{equation}
which satisfy $\gamma_{-n} = \gamma_n^*$ and read
\begin{equation}\label{gamma_transport_relation}
\gamma_0 = 1, \quad \gamma_{2n+1} = i \frac{\pi}{4}r^{2n}\alpha_{2n+1}, \quad \gamma_{2n} = - \frac{1}{(4 n^2 -1)} - \frac{\pi}{4}r^{2n-1}\beta_{2n}, \quad n\geq0.
\end{equation}    
Our focus will be on the following Toeplitz matrices, 
\begin{equation}
(T_N)_{ij} \equiv \gamma_{j-i}, \quad i,j = 0,1,\ldots,N 
\end{equation}
The reason is that $\{\gamma_n\}_{n=0}^N$ is a sequence of moments if and only if $T_N$ is positive semi-definite, $T_N \succeq 0$ (see Ref.~\cite{schmudgen2017moment}). We now discuss the consequences of this theorem for the transport coefficients in the $N=1,2,3$ cases. 

\paragraph{\underline{$N=1$ case:}} The Toeplitz matrix is given by 
\begin{equation}
T_1 = \begin{pmatrix}
1 & i\frac{\pi}{4}\alpha_1 \\
-i\frac{\pi}{4}\alpha_1 & 1 
\end{pmatrix}. 
\end{equation}
This matrix is positive semi-definite provided that $\det T_1 \geq 0$. This condition results in the two-sided bound $|\alpha_1| \leq \frac{4}{\pi}$, which is less sharp than Eq.~\eqref{a1_bound} and therefore superseded by it. 

\paragraph{\underline{$N=2$ case:}} The Toeplitz matrix is given by 
\begin{equation}
T_2 = \begin{pmatrix}
1 & i\frac{\pi}{4}\alpha_1 & -\frac{1}{3}-\frac{\pi}{4}r\beta_2 \\
-i\frac{\pi}{4}\alpha_1 & 1 & i\frac{\pi}{4}\alpha_1\\
-\frac{1}{3}-\frac{\pi}{4}r\beta_2 & -i\frac{\pi}{4}\alpha_1 & 1
\end{pmatrix}. 
\end{equation}
Together with Eqns.~\eqref{diffusion_other}, the requirement that $T_2 \succeq 0$ gives rise to the two-sided inequality \eqref{sound2}. 

In the case where $R\beta_2$ saturates the lower bound in Eq.~\ref{sound2}, all the higher-order transport coefficients are also fixed uniquely in terms of $\alpha_1$. The associated dispersion relation features poles and is therefore excluded. This shows that the curve $R\beta_2 = -\frac{16}{3\pi} + \frac{\pi}{2}\alpha_1^2$, $|\alpha_1|\leq 1$ does not belong to the hydrohedron. We refer the reader to Appendix \ref{app:edges} for additional details.

\paragraph{\underline{$N=3$ case:}} The positive semi-definiteness of $T_3$ 
leads to the two-sided bound \eqref{sound3}.  

As it happened in the $N=2$ case, the loci where any side of the bound \eqref{sound3} is saturated are outside the hydrohedron. The reason is as before: at these boundaries, $R^2 \alpha_3$ and all the higher-order transport coefficients are fixed uniquely in terms of $\alpha_1$ and $R\beta_2$, leading to a dispersion relation that features poles and is thus in conflict with the causality condition \eqref{cond}.

\section{Further details on the hydrohedron boundary}\label{app:edges}

\subsection{Diffusion}

In this Appendix, we explain why the boundaries of the diffusion cross-section of the hydrohedron determined by the moment problem are open. We emphasise that these boundaries have a different status than the ones associated to the condition $R\beta_2 \leq 0$; in principle, this second kind of boundary can contain dispersion relations that uphold \eqref{cond} and thus belong to the hydrohedron. The analysis that follows relies essentially on standard results in the moment problem literature \cite{schmudgen2017moment, schmudgen2020lectures}.

\noindent We start by defining the moment cone $\mathcal{S}_{m+1}$ as the set of all truncated moment sequences of length $m+1$ of all Radon measures in $[0,1]$ (the latter set being denoted as $M_+([0,1])$), i.e., 
\begin{equation}
\mathcal{S}_{m+1} \equiv \left\{s=(a_0,a_1,\ldots,a_m):a_j=\int_0^1 x^j d\mu(x), \,j=0,\ldots,m,\, \mu \in M_+([0,1])\right\},      
\end{equation}
and point out that every nontrivial moment sequence $s \in \mathcal{S}_{m+1}$ can be represented by a measure of the form
\begin{equation}
\mu = \sum_{j=1}^p m_j \delta_{x_j},~~~p \leq m+1,      
\end{equation}
with pairwise distinct roots $x_j \in [0,1]$, weights $m_j > 0$ for all $j$, and $\delta_{x_j}$ a Dirac measure at the point $x_j$. A central quantity in our analysis is the index of this representing measure, $\textrm{ind}(\mu)$, defined as  
\begin{equation}
\textrm{ind}(\mu) \equiv \sum_{j=1}^p \epsilon(x_j),~~~\textrm{where}~~~\epsilon(0) = \epsilon(1) = 1~~~\textrm{and}~~~\epsilon(x)=2~~~\textrm{for}~~~x\in(0,1),   
\end{equation}
with $\textrm{ind}(s)$ denoting the minimal index of all representing measures of $s$.  

\noindent The crucial result for us is Theorem 10.7 in Ref.~\cite{schmudgen2017moment} (Theorem 3.5 in Ref.~\cite{schmudgen2020lectures}). Among others, in this theorem the following statements are shown to be equivalent: 
\begin{itemize}
\item[(i)] $s \in \partial \mathcal{S}_{m+1}$. 
\item[(ii)] $\textrm{ind}(s) \leq m$. 
\item[(iii)] The representing measure is unique.  
\end{itemize}
From this result, it follows that the moment sequences lying at the boundary of the moment cone are of the form 
\begin{equation}
a_0 = \sum_{j=1}^p m_j = 1, \quad a_n = \sum_{j=1}^p m_j x_j^n,~~~n>0.        
\end{equation}
From the expression above, the inversion formula \eqref{inversion_formula} entails that 
\begin{equation}\label{w_boundary_master}
\ww(\kk) = -\frac{2i}{\pi}\left(1+\left(\kk - \frac{1}{\kk}\right)\arctanh{\kk} - \sum_{j=1}^p m_j \frac{2\kk^2(1-2x_j + \kk^2)}{(1+\kk^2)^2 - 4 x_j \kk^2}  \right).   
\end{equation}
These dispersion relations are associated to the boundaries of the moment cones and uniquely determined by the relevant set of weights $m_j$ and roots $x_j$. They feature poles at locations fixed by $x_j$; hence, as shown in Ref.~\cite{Heller:2022ejw}, they do not respect the causality condition \eqref{cond} everywhere in the complex $k$-plane and thus do not belong to the hydrohedron. This shows the main result of this subsection - that the boundaries of the diffusion cross-section of hydrohedron determined by the moment problem are open. We reiterate that this analysis does not apply to the other boundaries, prescribed by the $R\beta_2 \leq 0$ condition. 

With the main result established, we now specialize the general discussion above to the cases $m=1,2,3$ for the interested reader.

\paragraph{\underline{$m=1$ case:}} We have that $\textrm{ind}(s) = 1$, and hence there is a single root $x_1 \in \{0, 1\}$ and weight $m_1=1$. The case $x_1 = 0$ leads to $R\beta_2 = \frac{8}{3\pi}$ and is unphysical. We are left with the case where $x_1 = 1$, resulting in a moment sequence of the form 
\begin{equation}
a_{n\geq0} = 1. 
\end{equation}
Upon using Eq.~\eqref{inversion_formula}, one finds that this boundary point saturates all the lower bounds on $c_{2n}$ put forward in Ref.~\cite{Heller:2022ejw}, when specialised to a purely diffusive mode. 

\paragraph{\underline{$m=2$ case:}} The new cases have $\textrm{ind}(s) = 2$, and correspond to: 
\begin{itemize}
\item[(a)] $p=1$, $m_1=1$ and $x_1 \equiv x \in (0,1)$ but otherwise free, and
\item[(b)] $p=2$, $x_1=0$, $x_2=1$, $m_1=1-\alpha$, and $m_2 = \alpha$ with $\alpha \in (0,1)$.  
\end{itemize}
In case (a), 
the moment sequence takes the form 
\begin{equation}
a_{n\geq0} = x^n.   
\end{equation}
With the help of Eq.~\eqref{moment_formula}, one can easily show that this form implies that $R^3 \beta_4$ saturates the upper bound in the first inequality of Eq.~\eqref{diffusion2}, with $x$ a parameter labelling points on this boundary.

\noindent In case (b), the moment sequence takes the form 
\begin{equation}
a_0=1, \quad a_{n>0} = \alpha.      
\end{equation}
It can be readily checked that the relations above imply that $R^3 \beta_4 =-\frac{64}{15\pi}$, such that the lower bound in the first inequality of Eq.~\eqref{diffusion2} is saturated, with $\alpha$ labelling points on this boundary. The choice of weights and roots for case (b) gives back the dispersion relation \eqref{diffusion_linear_boundary} in Appendix \ref{app:diffusion} upon usage of Eq.~\eqref{w_boundary_master}.

\noindent We conclude by pointing out that there is a fundamental difference between the dispersion relations associated to cases (a) and (b). While both feature poles and hence violate \eqref{cond} somewhere in the complex $k$-plane, the region where \eqref{cond} is violated includes part of the unit disk for the former, but not for the latter. Note that this also means the transport coefficients from (a) do not form part of the closure of the hydrohedron, whilst transport coefficients from (b) do.

\paragraph{\underline{$m=3$ case:}} The new cases have $\textrm{ind}(s) = 3$. They are: 
\begin{itemize}
\item[(c)] $p=2$, $x_1=0$, $x_2 = x \in (0,1)$, $m_1=1-\alpha$, $m_2 = \alpha$ with $\alpha \in (0,1)$, and  
\item[(d)] $p=2$, $x_1=x \in (0,1)$, $x_2=1$, $m_1=\alpha$, $m_2=1-\alpha$ with $\alpha \in (0,1)$. 
\end{itemize}
For case (c), 
the moment sequence is given by 
\begin{equation}
a_0=1,\quad a_{n>0} = \alpha x^n.    
\end{equation}
This relation, together with Eq.~\eqref{moment_formula}, implies that $R^5 \beta_6$ saturates the upper bound given in Eq.~\eqref{diffusion3}. For case (d), 
\begin{equation}
a_{n \geq 0} = \alpha x^n + (1-\alpha),      
\end{equation}
from which it follows that $R^5 \beta_6$ saturates the lower bound given in Eq.~\eqref{diffusion3}. The dispersion relations associated with cases (c) and (d) can be readily found using Eq.~\eqref{w_boundary_master}. Both feature poles and are therefore discarded, implying that the boundaries of the hydrohedron set by Eq.~\eqref{diffusion3} are open. 

\subsection{Sound}

As discussed in Appendix \ref{app:sound}, the relevant moment problem for the sound geometry is a trigonometric one. In this case, the mathematical results necessary for our analysis can be found in Chapter 11 of Ref.~\cite{schmudgen2017moment}. 

For the truncated trigonometric moment problem, the moment cone $\mathcal{S}_{m+1}$ is defined as the set of all moment sequences of length $n+1$ associated to all Radon measures on the unit circle $\mathbb T$ (the latter set being denoted as $M_+(\mathbb T)$), 
\begin{equation}
\mathcal{S}_{m+1} \equiv \left\{s=(\gamma_0, \gamma_1,\ldots,\gamma_n): \gamma_j = \int_{\mathbb T} z^{-j}d\mu(z),\, \mu\in M_+(\mathbb T) \right\}.     
\end{equation}
A sequence $s$ belonging to the boundary of $\mathcal{S}_{m+1}$ has a unique representing measure supported on at most $m$ points. From this result, it follows that if $s \in \partial \mathcal{S}_{m+1}$ then 
\begin{equation}
\gamma_n = \sum_{j=1}^p m_j e^{-i n \theta_j},\quad \theta_j \in [0,2\pi)    
\end{equation}
and $p\leq m$. We now explore the consequences of this assertion for the cases $m=2,3$. We do not consider the $m=1$ case since it is beyond the stiff fluid facets we discuss in Appendix \ref{app:perfectfluid}.

\paragraph{\underline{$m=2$ case:}} The moment sequence is of the form 
\begin{equation}\label{gamma_sequence_m=2}
\gamma_n = \frac{1}{2} \left(e^{-i n \theta} + e^{-i n (\pi-\theta)} \right).   
\end{equation}
This expression implies that $R \beta_2$ saturates the lower bound in Eq.~\eqref{sound2}. Upon using \eqref{gamma_transport_relation} and \eqref{omega_series}, \eqref{gamma_sequence_m=2} results in the dispersion relation 
\begin{equation}\label{w_12_sound}
\ww(\kk) = -\frac{2i}{\pi}\left((\kk-\kk^{-1})\arctanh(\kk) + \frac{1-\kk^4}{1-2\cos(2\theta)\kk^2+\kk^4}\right) - \frac{4\sin(\theta)}{\pi} \frac{\kk (1+\kk^2)}{1-2 \cos(2\theta)\kk^2+\kk^4},     
\end{equation}
with $\sin(\theta) = -\frac{\pi}{4}\alpha_1$. This dispersion relation does not belong to the hydrohedron: it features poles and moreover it does not belong to its closure, since it also violates \eqref{cond} in the vicinity of $k^2=1$ inside the unit disk. 

\paragraph{\underline{$m=3$ case:}} The representing measure has three support points. There are two candidate moment sequences such that $\gamma_{2n+1}$ is purely imaginary and $\gamma_{2n}$ purely real, 
\begin{equation}\label{gamma_sequence_m=3}
\gamma^{(\pm)}_n = \alpha \left(e^{-i n \theta} + e^{-i n (\pi-\theta)}\right) + (1-2\alpha)e^{\pm\frac{i n \pi}{2}}.      
\end{equation}
For $\gamma_n^{(+)}$, $R^2 \alpha_3$ saturates the lower bound in Eq.~\eqref{sound3}; for $\gamma_n^{(-)}$, it saturates the upper bound on the same equation. The dispersion relations associated with \eqref{gamma_sequence_m=3} are
\begin{equation}
\begin{split}
&\ww^{(+)}(\kk) = -\frac{2i}{\pi}\left((\kk-\kk^{-1})\arctanh{\kk} + \frac{(1-\kk^2)(1-2 (\cos(2\theta)-4\alpha\cos(\theta)^2)\kk^2 + \kk^4)}{(1+\kk^2)(1-2\cos(2\theta)\kk^2 + \kk^4)} \right)+ \\
&\frac{4}{\pi}\kk \frac{1-2\alpha}{1+\kk^2}-\frac{8}{\pi}\kk \frac{\alpha \sin(\theta) (1+\kk^2)}{1-2\cos(2\theta)\kk^2+\kk^4},  
\end{split}
\end{equation}
\begin{equation}
\begin{split}
&\ww^{(-)}(\kk) = -\frac{2i}{\pi}\left( (\kk-\kk^{-1})\arctanh{\kk} + \frac{(1-\kk^2)(1-2(\cos(2\theta)-4\alpha \cos(\theta)^2)\kk^2 + \kk^4)}{(1+\kk^2)(1-2 \cos(2\theta)\kk^2 + \kk^4)} \right) - \\
& \frac{4}{\pi} \kk \frac{1-2\alpha}{1+\kk^2} - \frac{8}{\pi}\kk\frac{\alpha \sin(\theta) (1+\kk^2)}{1-2\cos(2\theta)\kk^2 + \kk^4}.   
\end{split}
\end{equation}
As mentioned in the main text, both feature poles and are in conflict with the causality condition \eqref{cond}. 

\section{The stiff fluid facets\label{app:perfectfluid}}
Given the sound mode dispersion relation \eqref{omegasound} with luminal sound speed $\alpha_1 = \pm 1$, the causality condition \eqref{cond} uniquely fixes the dispersion relation to be $\omega(k) = \pm k$. To prove this, first note that
\be
|\im\,k| - \im\,\omega(k) = |r \sin\theta| - \sum_{n=0}^\infty \alpha_{2n+1} r^{2n+1} \sin\left((2n+1)\theta\right) - \sum_{n=1}^\infty \beta_{2n} r^{2n} \cos\left(2n\theta\right).
\ee
By \eqref{cond}, this must be non-negative for all $0\leq r<R$ and $0 \leq \theta < 2\pi$ (for $r\geq R$ the Taylor series representation is invalid).
Taking $\alpha_1 = 1$ and restricting to $\theta \in [0,\pi]$ gives a cancellation between the first two terms in this $r$ expansion,
\be
|\im\,k| - \im\,\omega(k) =  -\sum_{n=1}^\infty \alpha_{2n+1} r^{2n+1} \sin\left((2n+1)\theta\right) - \sum_{n=1}^\infty \beta_{2n} r^{2n} \cos\left(2n\theta\right).
\ee
Thus the leading term in this expansion is now at order $r^2$.
Let us proceed by induction. Take all coefficients up to order $r^{m-1}$ to be zero, with the exception of $\alpha_1$, and note that $m\geq 2$. If $m$ is even, then the next term in the expansion gives the constraint (in the limit $r\to 0$)
\be
-\beta_{m/2} \cos(m\theta) \geq 0, \qquad \forall \theta \in [0,\pi]
\ee
and hence $\beta_{m/2} = 0$. If instead $m$ is odd, then the next term in the expansion gives the constraint (in the limit $r\to 0$)
\be
-\alpha_{(m-1)/2} \sin(m\theta) \geq 0, \qquad \forall \theta \in [0,\pi]
\ee
and thus similarly $\alpha_{(m-1)/2} = 0$. Hence all coefficients zero up to order $r^{m-1}$ implies the coefficients at order $r^m$ are zero.
Finally we note that the base case $m=2$ is covered by the above analysis, $\beta_2 = 0$, which completes the proof for $\alpha_1 = 1$. The proof for $\alpha_1 = -1$ proceeds analogously. We note the observation that $\alpha_1 = \pm 1 \implies \beta_2 = 0$ was given also in \cite{Gavassino:2023myj}.

\bibliographystyle{JHEP}
\bibliography{causality} 

\end{document}